# Frequency-Resolved Spatial Beam Mapping in Multimode Fibers: Application to Mid-Infrared Supercontinuum Generation

Y. LEVENTOUX,[1] G. GRANGER,[1] K. KRUPA,[2] T. MANSURYAN,[1] M. FABERT,[1] A. TONELLO,[1] S. WABNITZ,[3] V. COUDERC[1] AND S. FEVRIER[1,*]

[1] *Université de Limoges, XLIM, UMR CNRS 7252, 123 Av. A. Thomas, 87060 Limoges, France*
[2] *Institute of Physical Chemistry, Polish Academy of Sciences, Warsaw, ul. Kasprzaka 44/52, 01-224 Poland*
[3] *Dipartimento di Ingegneria dell'Informazione, Elettronica e Telecomunicazioni, Sapienza University of Rome, Via Eudossiana 18, 00184 Rome, Italy*
*Corresponding author: sebastien.fevrier@unilim.fr

**We present a new spatial-spectral mapping technique permitting to measure the beam intensity at the output of a graded-index (GRIN) multimode fiber with sub-nanometric spectral resolution. We apply this method to visualize the fine structure of the beam shape of a sideband generated at 1870 nm by geometric parametric instability (GPI) in a GRIN fiber. After spatial-spectral characterization, we amplify the GPI sideband with a Tm-doped fiber amplifier to obtain a microjoule-scale picosecond pump whose spectrum is finally broadened in a segment of InF$_3$ optical fiber to achieve supercontinuum ranging from 1.7 μm up to 3.4 μm.**

Fiber laser technology and more specifically all-fiber based light sources have made huge progress over the past few decades. Nowadays, supercontinuum (SC) laser sources based on silica core optical fibers are commercially available, spanning from the visible to the near-infrared spectral region. However, the spectral range of short-wave (SW) and middle-wave infrared (MID-IR) remain yet poorly covered by current fiber-based solutions although these spectral ranges are crucial for numerous important applications, such as for medicine, spectroscopy, imaging, material processing, and remote environmental sensing. Indeed, several gas molecules, such as water vapor, carbon dioxide ($CO_2$), and nitrous oxide ($N_2O$) exhibit strong absorption lines in the SW/MID-IR spectral ranges, ranging from 2 μm to approximately 12 μm light from such spectral regions is also strongly absorbed by biological tissues, with direct applications in dermatology and more generally, in surgery. Among the different approaches to cover SW/MID-IR wavelengths, Thulium and Thulium-Holmium co-doped fiber lasers have been considered as platforms for building ultrafast pulse sources around the wavelength range of 2 μm [1]. Tm-doped mode-locked fiber lasers have been already exploited for SW/MID-IR SC generation in non-silica optical fibers including, for instance, tellurite and chalcogenide fibers [2,3]. The advent of gas-filled hollow core photonic crystal fibers (HC-PCFs) has also opened new directions for the generation of SW/MID-IR broadband sources via stimulated Raman scattering (SRS) [4]. Fluoroindate fibers pumped by an amplified diode at 1.5μm have been considered for MID-IR SC generation [5]; Z. Eslami *et al.* [6] studied SC generation in multimode chalcogenide fiber, pumped with a femtosecond optical parametric amplifier. A. Parriaux *et al.* reported a SW/MID-IR all-fibered dual-comb spectroscopy source by using fourth-order modulation instability [7].

Over the last years, multimode optical fibers (MMFs) have also attracted a great attention [8]. Recent studies demonstrated spatial beam self-cleaning in MMFs [9,10], mode-locking by spatiotemporal filtering [11], or the generation of parametric sidebands by geometrical parametric instabilities (GPI) [12,13]. Applications to endoscopy [14,15], nonlinear imaging [16], or laser sources [17] are currently conceived; new measurement techniques have also been developed, in order to capture the spatiotemporal nature of the light propagation [18-20].

In this Letter, we exploit the GPI to develop a new all-fiber based SC laser source emitting high-energy light in the SW/MID-IR starting from conventional laser sources at 1064 nm. The beam profile needs to be measured at all wavelengths: for this reason, we develop here an *ad-hoc* measurement technique. Specifically, we demonstrate how the GPI Stokes sideband located at 1870 nm can serve as an efficient pump for SW/MID-IR SC generation. In order to analyze the fine spatial content of the sideband at 1870 nm, we have modified our 3D mapping technique for multimode beams, initially developed for spatio-temporal direct detection [21], by replacing the ultrafast oscilloscope with an optical spectrum analyzer. In doing so, we are able to visualize the beam profile for each spectral component of the guided light with sub-nanometric spectral resolution. Specifically, this spatio-spectral (SS) technique, which stems from the spatial sampling method [22], permits us to capture the fine spatial structure of the GPI peaks, showing how the spectrum is locally composed by several multimode spectral components. A similar 3D map for time domain was proposed in [23]. GPI is a collective spatiotemporal nonlinear instability, which originates from the four-wave mixing of many-mode nonlinear interactions in a GRIN fiber. Overall, the guided light undergoes periodical longitudinal oscillations, able to induce temporal instabilities in the presence of dispersion and the Kerr effect. As a result, it is possible to generate a strong pair of resonant peaks with extremely large frequency detuning even with a pump beam propagating in the normal dispersion regime, which opens interesting applications based on the frequency conversion of powerful lasers at 1 μm. As a result, with this technique it is possible to frequency down-convert a laser pump nearly to the upper wavelength limit of conventional silica glass fibers, right before the strong attenuation in the SW infrared. Indeed, the frequency detuning of the first Stokes (or anti-Stokes) sidebands from the pump is approximatively $f_1 = \sqrt{1/(2\pi\Lambda\kappa'')}$, where $\Lambda = \pi R/\sqrt{2\Delta}$ is the fiber self-imaging period for symmetric beams, $R$ is the fiber core radius, $\Delta$ is the relative refractive index difference, and $\kappa''$ is the group velocity dispersion at the pump wavelength. This simple formula can guide the choice of a standard commercially available GRIN fiber, whose GPI Stokes sideband falls within the range of amplification of Thulium-doped fibers when the GRIN fiber is pumped by a 1064 nm laser. To overcome the strong absorption of silica fiber in SW/MID-IR, our all-fiber system is then composed of three stages, as illustrated in Fig.1(a). In the first stage, a segment of GRIN fiber generates a GPI sideband at 1870 nm. In the second stage, a Tm-doped fiber amplifier, pumped by an Erbium-Doped Fiber Laser (EDFL) boosts the power of such narrow sideband and finally, with the help of a short segment of single-mode fiber at 2μm (SMF 2000), the SW/MID-IR spectral broadening takes place in the third stage, in a fluoroindate glass ($InF_3$) fiber.

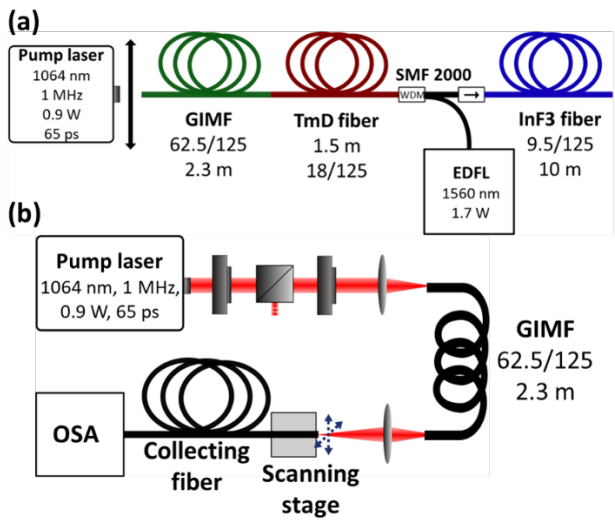

Fig. 1. (a) Experimental set-up for GPI-based SW/MID-IR SC generation. (b) Set-up for spatial-spectral beam mapping.

The first stage of our system is composed by a standard 2.3 m-long GRIN fiber with 62.5 μm core diameter and refractive index difference Δn = 0.027 (numerical aperture NA=0.275 ). We pumped the fiber with a laser at 1064 nm, delivering 65 ps pulses (measured at $1/e^2$ of their maximum) with a repetition rate of 1 MHz. The linearly polarized Gaussian beam was focused on the input face of the fiber with a diameter of 20 μm (measured at $1/e^2$). To measure the output optical spectra, we used two optical spectrum analyzers (OSAs) to cover all the relevant spectral range from 500 to 3400 nm.

We applied our 3D mapping technique to the first stage: the structure of the set-up is schematically illustrated in Fig.1(b). In particular, the magnified image at the output of the GRIN fiber is scanned by moving a single-mode fiber (SMF) in the transverse plane. At each position of the SMF, the corresponding spectrum which is detected by the OSA is stored in a 3D matrix, jointly with the corresponding coordinates of the SMF in the transverse plane. Note that, although the 3D matrix is composed by spectral measurements obtained from different laser pulses, the stability of the setup permits the post-processing of the 3D matrix in order to reconstruct the beam image at each accessible wavelength. Specifically, this technique permits a spectral resolution of 0.1 nm (only limited by the resolution of the OSA). Note that such value of resolution can be hardly obtained with an optical filter; moreover, the present technique permits to reconstruct the spatial profile along the entire spectral window, and with the spectral resolution of the OSA. The critical point is certainly the stability from shot to shot of the beam pattern along the measurement process. We notice that the spatial profiles of the various modes excited at a fixed wavelength are not directly detected as it is the case with a standard camera: the image is the result of the application of a post-processing routine starting from the spectra collected by long series of measurements with the OSA. Direct comparisons with an optical filter and a standard camera confirmed the applicability of this technique.

Fig. 2 illustrates a typical outcome of our measurement technique. Panel (a) shows the dependence of spatial beam profiles upon wavelength around the region of the first GPI Stokes sideband at 1870 nm. Fig. 2(b) illustrates the experimental spectrum recorded at the output of the GRIN fiber (first stage) at the pump power of 900 mW, where it is possible to identify the pair of GPI sidebands. The corresponding spatial-spectral analysis for the GPI anti-Stokes sideband at 732 nm is illustrated in panel (c). Figure 2(c) shows that spectral components constituting the anti-Stokes sideband turn out to be mainly carried by the fundamental mode of the fiber. This observation is in good agreement with previous observations, which however were averaged over the 10 nm range of bandpass optical filters [13]. On the other hand, Figure 2(a) shows that the Stokes peak is mainly carried by higher-order guided modes (the output intensity profiles indicate the presence of $LP_{11}$, $LP_{02}$, $LP_{03}$, $LP_{04}$ modes): we interpret the lack of azimuthal symmetry in the observed beam shapes as a result of the cumulated effect of random fiber deformations. We can also notice how the GPI Stokes sideband has a fine structure with multiple humps of lower intensities, and whose beam shape is fully accessible with the present measurement technique. The beam shapes around 1870 nm are suitable for being amplified in a Tm-doped fiber amplifier, and to be transformed into a single-mode SW/MID-IR SC by the last two stages.

In order to simulate the most salient features of the experimental spectrum from the first stage, we considered a GRIN fiber with R = 32.5 μm, Δ = 0.0172, Δn = 0.0257, NA=0.275. The model is based on a scalar version of the 3D+1 generalized nonlinear Schrödinger equation (GNLSE), comprising diffraction, waveguide contribution and frequency dependence of the refractive index and losses, Raman and Kerr effect. A further coarse step is applied to include the effect of fiber disorder. More specifically the core shape is deformed to an ellipse with random orientation (uniformly distributed) and random axes variation of 0.2 μm every 10 mm. This method permits to consider, at least in a qualitative manner, the physical fiber deformations and stress at the base of the speckle formation along the propagation. An example of our results is illustrated in Fig. 3 for an input intensity of 40 GW/cm² and a pulse duration of 16 ps. The input intensity was increased with respect to the experiment in order to reduce the propagation length and so the computational time. In particular, Fig. 3(a) shows the spectral output of the GPI sideband. The three panels illustrate sample cases of spatial beam shape obtained at the corresponding spectral position identified by the purple dashed lined. Panel (b) of the same figure illustrates the build-up of the sidebands along the fiber. With a graphical superposition, we also report the phase mismatch Δk among Stokes (S), anti-Stokes (AS) and pump (P), by including the frequency variation of the refractive index n(ω).

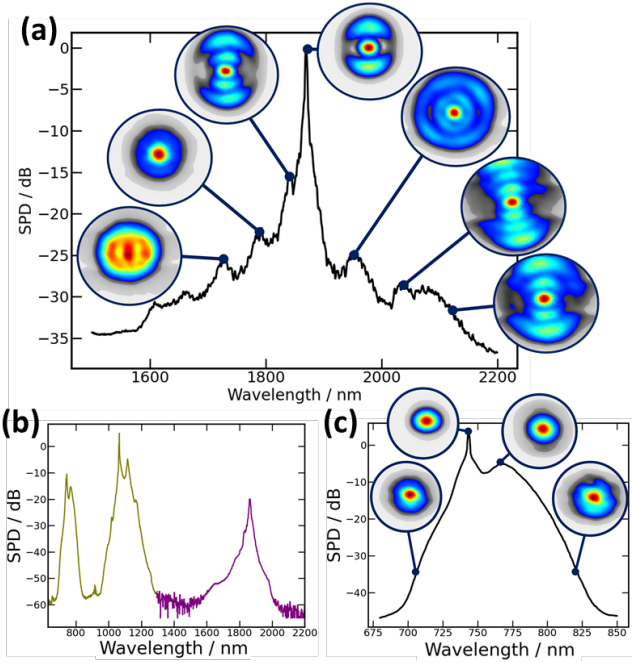

fiber permits a selective coupling in order to generate a SC covering the spectral range up to 3400 nm with a maximum output power of 75 mW.

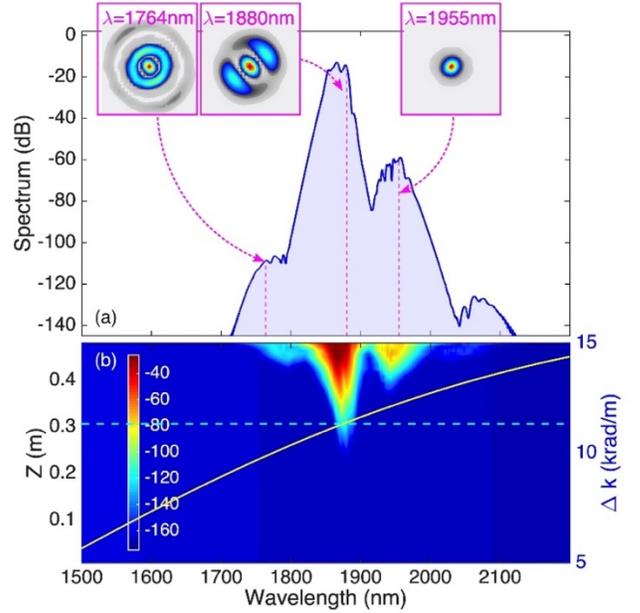

Fig. 3. (a) Numerical spectrum in Z = 0.48 m. The insets show the corresponding spatial profile at the selected spectral positions (dashed lines). (b) Numerical propagation upon fiber length; phase-matching when Δk (yellow curve) crosses the momentum 2π/Λ (dashed azure line).

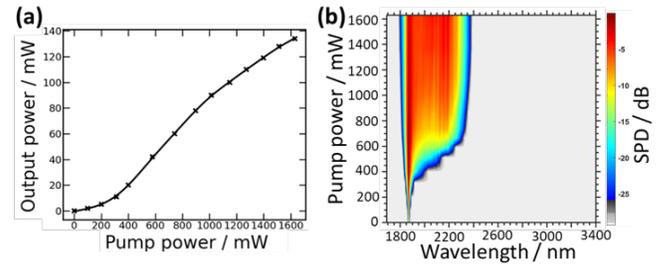

Fig. 2. (a) 2D reconstructed representation of the spatial output pattern at selected spectral components around the Stokes GPI at 1870 nm. (b) Experimental spectrum at the output of the GRIN fiber (900 mW laser power). (c) Same representation as in (a) but referred to the GPI anti-Stokes.

Neglecting the nonlinear contributions to the wave-vectors, one has Δk = k(ω$_S$) + k(ω$_{AS}$) - 2k(ω$_P$), where k(ω) = ωn(ω)/c. The phase matching is graphically identified by the crossing point Δk = 2π/Λ; its position agrees well with the full numerical solution of the 3D+1 GNLSE. It is interesting to observe that the GPI sideband in the simulation is not composed by a unique sharp peak. Besides the dominant peak, there is a satellite peak with a different spatial content. These full numerical simulations permit to confirm the coexistence of different spatial profiles. The lack of azimuthal symmetry in the output beams is the result of multiple deformations applied in the coarse step method. With weaker deformations (e.g. random axes variation of 0.1 μm or less, here not shown) the pattern was composed by concentric rings. The conversion efficiency and bandwidth at 1870nm were larger than in the experiments. The qualitative agreement of the simulations in Fig.3 with the experiments of Fig. 2 let then suppose that the fiber disorder contributes by reducing the sideband width and the conversion efficiency around 1870 nm as well as by breaking the azimuthal symmetry of pattern into lobes.

In the second and third stage of our all-fiber SC source, we exploited the previously induced GPI Stokes sideband to build a 1870 nm laser pump for the SW/MID-IR SC generation. As it is schematically presented in panel (a) of Fig. 1, we used a 1.5 m long 18/125 Tm-doped fiber, which was backward-pumped by an EDFL at 1560 nm. Figure 4 reports the results of the amplification step by illustrating in panel (a) the output power measured as a function of the EDFL-pump power and in panel (b) the corresponding evolution of the output spectral profile.

Finally, the sideband at 1870 nm is injected in the 10 m-long 9.5/125 InF$_3$ fiber by using a 1-m long segment of SMF-2000. Such single-mode

Fig. 4. (a) Power measured at the output of Tm-doped fiber amplifier as a function of EDFL pump power. (b) Corresponding evolution of the spectral profile. GPI Stokes sideband generated at 900 mW of 1064 nm laser power was used as an input signal.

The InF$_3$ fiber (9.5/125 from Le Verre Fluoré) has a cut-off wavelength at 3.7 μm, and NA=0.3. The group velocity dispersion and third order dispersion for InF$_3$ are $β_2$=-1.51×10$^{-26}$ s$^2$/m and $β_3$=1.39×10$^{-40}$ s$^3$/m (the values refer to the material dispersion). The resulting spectral evolution and SC output power versus EDFL-pump power are shown in panel (b) and (a) of Fig.5, respectively. To better appreciate the whole process of the SW/MID-IR SC generation, we summarize in Fig. 6 the spectra obtained after each step of the all-fiber system, including the spectrum of the first GPI Stokes sideband emerging at the output of GIMF (blue curve), the spectrum at the output of Tm-doped fiber amplifier (TDFA) (green curve), and the spectrum at the output of the InF$_3$ fiber (red curve); the GPI spectrum was induced at the 1064 nm pump power of 900 mW, while the last two spectra were generated at the maximum EDFL pump power of 75 mW.

To verify the spatial features of the obtained energetic SW/MID-IR SC, we applied again our novel SS-technique, which demonstrates that the beam is carried by a single-mode across the whole IR spectrum (see the inset of Fig.6).

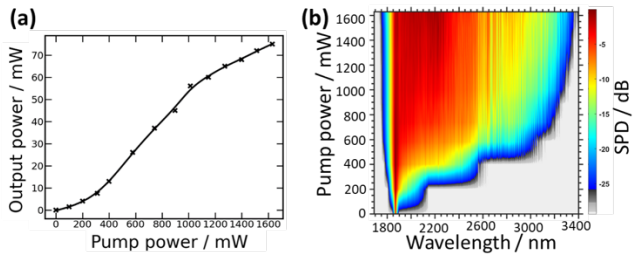

Fig. 5. (a) Power measured at the output of InF$_3$ fiber as a function of EDFL pump power. (b) Corresponding evolution of the spectral profile.

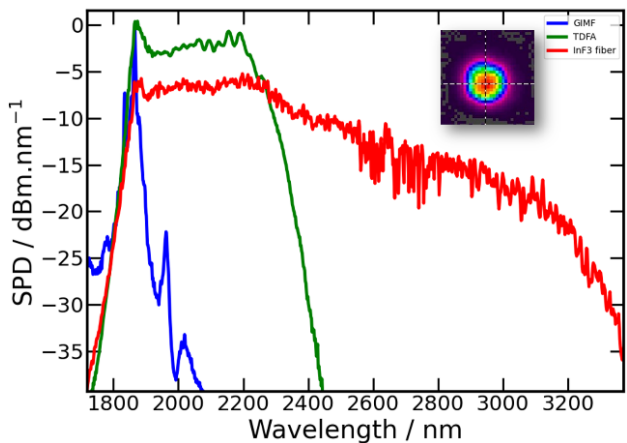

Fig. 6. Comparison of the experimental spectra obtained at the output of GIMF including only the Stokes GPI sideband (blue curve), the output of TDFA (green curve), and at the output of the InF$_3$ fiber (red curve). SPD: spectral power density. Insets: Output beam intensity patterns of the SC induced in InF$_3$ fiber.

In conclusion, we experimentally demonstrated a novel technique to obtain a high beam quality, spatially nearly single-mode SC laser source delivering 1 MHz pulses in the spectral range spanning from 1.7 µm up to 3.4 µm. MID-IR SC generation in InF$_3$ glass fiber was early reported in [5] with an amplified laser diode at 1.55 µm. Our work shows how MID-IR SC can be also generated by a novel high energy (~ 1 µJ) picosecond GPI-based laser source at 1870 nm, starting from a laser pump around 1 µm. In particular, we used a TDFA to amplify the GPI Stokes sideband induced in a standard 62.5/125 GIMF by a 65 ps 1MHz laser at 1064nm. The MID-IR SC was generated in a 10-m long InF$_3$ optical fiber. The fine tuning of this system stimulated the development of a new 3D beam diagnostics that allows for capturing a spectrally resolved intensity image of a beam with 0.1 nm resolution (as fixed by the resolution of the OSA). We underline that our method is not single-shot, and it relies on the stability of the beam shape from shot to shot. By using this technique, we demonstrated that the first pair of GPI sidebands, and in particular its Stokes peak, are composed by spectral elements carried by high-order guided modes. We believe that our beam diagnostic tool can have range of applications far larger than the example illustrated in the present work and can help to study the spatial features and modal composition of a laser beam at wavelengths which are otherwise not available by using standard optical filters. Thanks to the achievable high spectral resolution, this instrument can also open the way for further investigations, and ultimately lead to a better understanding of spatio-spectral dynamics of multimode beam propagation and laser sources.

**Funding.** Arianegroup (X-LAS); Direction Générale de l'Armement (iXcore); Conseil Régional Aquitaine (F2MH, FLOWA, SI2P); Agence Nationale de la Recherche (ANR-10-LABX-0074-01, ANR-15-IDEX-0003, ANR16-CE08-0031, ANR-18-CE080016-01); H2020 Marie Skłodowska-Curie Actions (713694); H2020 FET Open PETACom; European Research Council (740355, 874596).

**Disclosures.** The authors declare no conflicts of interest.

**Data Availability.** Data underlying the results presented in this paper are not publicly available at this time but may be obtained from the authors upon reasonable request.